# Application Research of Spline Interpolation and ARIMA in the Field of Stock Market Forecasting

Xitai Yu

Joint college of science, Shandong University, Shandong Weihai 264209, China
202100700065@mail.sdu.edu.cn

**Abstract.** Time series forecasting has always been a widely applied field, contributing to the development of various domains such as business, finance, operations, supply chain management, risk management, and scientific research. The ARIMA (Autoregressive Integrated Moving Average model) has extensive applications in the field of time series forecasting. However, the predictive performance of the ARIMA model is limited when dealing with data gaps or significant noise. Spline interpolation is a numerical analysis method used to construct a smooth interpolation function given a set of data points, facilitating interpolation and extrapolation between these data points. Based on previous research, we have found that cubic spline interpolation performs well in capturing the smooth changes of stock price curves, especially when the market trends are relatively stable. Therefore, this paper integrates the two approaches by taking the time series data of seven dimensions in stock trading – namely, opening price, highest price, lowest price, closing price, trading volume, dividends, and stock splits – as an example. It explores the application of cubic spline interpolation and ARIMA in the field of time series forecasting. Furthermore, it establishes a time series forecasting model based on cubic spline interpolation and ARIMA. Through validation, the model has demonstrated certain guidance and reference value for short-term time series forecasting.

## 1 Introduction

Accurate time series forecasting in fields such as business, finance, and operations can assist managers in making more informed decisions, such as production planning, inventory management, and marketing strategies. Hence, accurate time series forecasting has significant implications for decision-making, planning, risk management, and investment across various domains. Accurate forecasting of time series remains a pressing issue in need of resolution. ARIMA model is one of the most classical methods for time series forecasting. In the context of urban rail transit, ARIMA can be utilized to accurately predict short-term passenger traffic, assisting stations in effectively managing peak passenger flows [1]. In the field of environmental protection, the ARIMA model can be applied to achieve short-term and accurate predictions of environmental indicators. These applications include forecasting chlorophyll-a concentration in reservoirs [2], predicting oceanic El Niño indices [3], and even global surface temperature predictions [4]. ARIMA still finds extensive applications in fields such as engineering construction, commerce, and healthcare, including short-term wind speed prediction [5], cross-border e-commerce trend analysis [6], analysis of tuberculosis epidemic trends [7], and so on. Lujun Xu and others [8], compared the predictive performance of the moving average method and the ARIMA model using corporate cash flow data as an example. They found that the ARIMA model had a higher prediction accuracy than the moving average method by 4.61%. This conclusion supports that the ARIMA model provides more accurate predictions. Gao Long [9] employed the TOPSIS method to select a material of interest for research. They then constructed a short-term material production demand forecasting model based on ARIMA. However, due to the limited time series data available, the predictive accuracy is constrained. Lijuan Wang and Hui Suo [10] investigated the application of cubic spline interpolation and machine learning in handling missing values in time series forecasting. They found that cubic spline interpolation can make data more continuous, evenly distributed, and trend-focused, ultimately enhancing the performance of time series forecasting models. This paper utilizes data from a certain stock spanning from 2010 to 2022, covering a total of 12 years, with daily measurements in seven dimensions: opening price, highest price, lowest price, closing price, trading volume, dividends, and stock splits. A time series forecasting model was established based on cubic spline interpolation and ARIMA. The model was trained and tested using a rolling window approach for step-by-step prediction.

## 2 Method and Experiment

### 2.1 Introduction to the Dataset

The dataset records indicators for seven aspects - 'Open', 'High', 'Low', 'Close', 'Volume', 'Dividends', and 'Stock Splits' - starting from January 4th, 2010, until December 3rd, 2022. Each of these indicators can be viewed as a time series. Each row represents a day, and each column represents a stock indicator. The dataset contains 3272 samples, which is significantly less than the number of days between January 4th, 2010, and December 3rd, 2022. This indicates the presence of missing values in the dataset. Due to space limitations, only the ARIMA forecasting model based on the 'Open' and 'Close' time series will be presented below.

### 2.2 Method

**Cubic Spline Interpolation.**
Using spline functions for interpolation is known as spline interpolation. If the spline function is a piecewise polynomial of degree 3, it is referred to as cubic spline interpolation. In other words, given a function $y = f(x)$ and the values of the function at n+1 nodes $y_i = f(x_i)(i = 0,1\cdots n)$ in the interval, compute the interpolation function

$S(x)$ such that $S(x)$ is a piecewise cubic function on $[a,b]$, and it has continuous second derivatives. From this, the expression for $S(x)$ can be derived along with the system of equations for solving the undetermined parameters:

$$S(x) = \{S_i(x) = a_i x^3 + b_i x^2 + c_i x + d_i, x \in [x_i, x_{i+1}], i = 0,1 \cdots n-1\}$$

st.

$$\begin{aligned}
&S(x_i) = y_i (i = 0,1 \cdots n) \\
&S_i(x_{i+1}) = S_{i+1}(x_i)(i = 0,1 \cdots n-2) \\
&S_i^{(1)}(x_{i+1}) = S_{i+1}^{(1)}(x_i)(i = 0,1 \cdots n-2) \\
&S_i^{(2)}(x_{i+1}) = S_{i+1}^{(2)}(x_i)(i = 0,1 \cdots n-2) \\
&S^{(1)}(a+0) = S^{(1)}(b-0) \\
&S^{(2)}(a+0) = S^{(2)}(b-0)
\end{aligned} \quad (1)$$

Utilizing cubic spline interpolation for preprocessing time series data can fill in missing data, resulting in denser and more continuous data, with a more even distribution and a more focused trend. It enhances the data's analytical and predictive capabilities.

**Autoregressive Integrated Moving Average Model**
The general form of an Autoregressive Moving Average Model (ARMA) model is:

$$X_t = \phi_0 + \phi_1 X_{t-1} + \cdots + \phi_p X_{t-p} + \varepsilon_t + \theta_1 \varepsilon_{t-1} + \cdots + \theta_q \varepsilon_{t-q} \quad (2)$$

Where $\{\varepsilon_t\}$ is a zero-mean independent identically distributed white noise with variance $\sigma^2$, and it is independent of $\{X_t\}$. $\{\phi_i\}_{i=0}^p$ and $\{\theta_i\}_{i=0}^q$ are the parameters to be fitted.

$$1 - \phi_1 z - \phi_2 z^2 \cdots + \phi_p z^p = 0 \quad (3)$$

$$1 - \theta_1 z - \theta_2 z^2 \cdots + \theta_q z^q = 0 \quad (4)$$

The stationarity condition requires that the roots $z_*$ of the characteristic equation (3) lie outside the unit circle, which means $|z_*| > 1$. The invertibility condition requires that the roots $z_*$ of the characteristic equation(4) also lie outside the unit circle, which means $|z_*| > 1$.

ARMA models are effective in handling stationary time series, but they can't directly address non-stationary time series with trends or seasonal components, such as random walks or trended random walks. In the former, the variance increases linearly with time, and in the latter, both the mean and variance increase linearly with time. Both are non-stationary time series, but their linear trends can be removed through differencing, transforming them into stationary and reversible time series.

Therefore, we can extend the application of ARMA models by introducing differencing to handle partially non-stationary time series, which gives rise to ARIMA models. In form, ARIMA models are similar to ARMA (p, q) models, but among their p+1 characteristic roots, one is equal to 1, while the remaining characteristic roots lie outside the unit circle. To model with an ARIMA model, you simply calculate the differences of the original non-stationary sequence, and once it becomes stationary, you can build an ARMA model on the differences.

**Modeling process**

First, apply cubic spline interpolation to the time series with missing values. Then conduct stationarity test on the resulting sequence. If the test does not pass, the sequence is differenced to enhance its stationarity. Once the stationarity test is successful, based on the stationary sequence, the model order is determined, followed by model training and evaluation in sequence. If the model does not pass, the above steps are repeated until the most suitable model order is selected. Finally, use the chosen model to forecast future data.

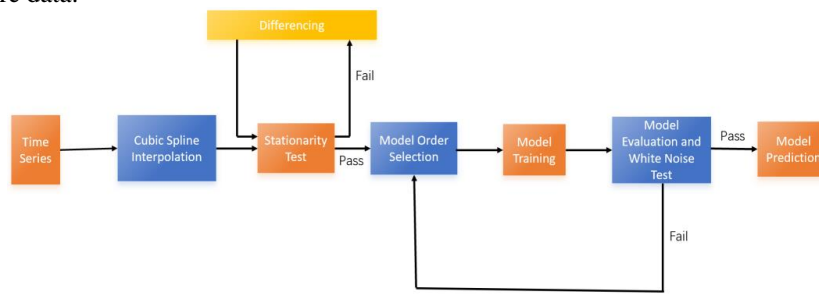

**Fig. 1.** A graph that shows Model Process. (Photo/Picture credit :Original）

### 2.3 Data Interpolation

The core assumption of the ARIMA model is the continuity and temporal dependency between data points, while missing values disrupt these assumptions and lead to decreased accuracy of the model. Through previous analysis, interpolation can fill in missing values, maintaining the continuity of the time series and better reflecting the actual temporal dependencies and patterns. This aids the model in capturing trends and seasonality, thus enhancing prediction accuracy. Therefore, before training the model, it's necessary to perform cubic spline interpolation on the time series .

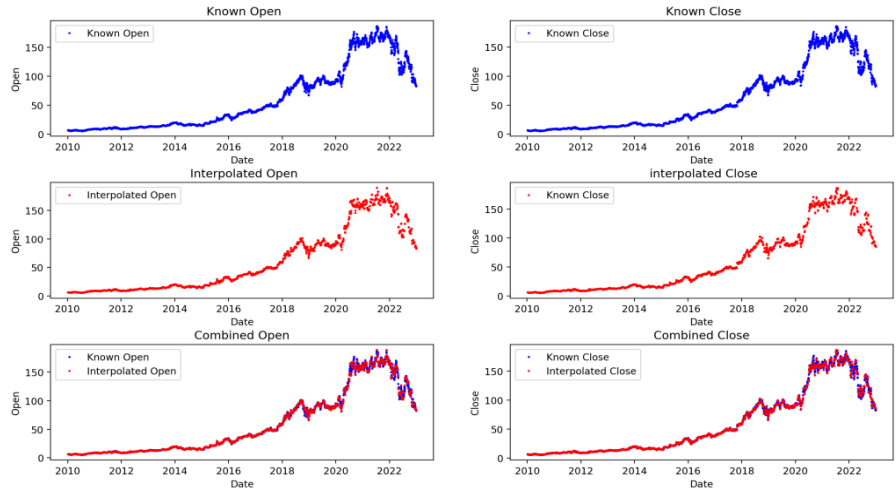

**Fig. 2.** Cubic spline interpolation results.(Photo/Picture credit :Original)

Fig.2 visualizes the interpolation results with 'Date' on the x-axis and either 'Open' or 'Close' on the y-axis. It's evident that the interpolated time series exhibits enhanced continuity and predictability.

### 2.4 Stationarity Test

Using the Augmented Dickey-Fuller (ADF) test for stationarity on the 'Open' and 'Close' series, the results are as Table1.

**Table 1.** Stationarity Test Results.

|  |  | Open | Open(first-order differenced) | Close | Close(first-order differenced) |
|---|---|---|---|---|---|
| ADF statistic | | -1.09 | -14.21 | -1.01 | -13.81 |
| p-value | | 0.71 | 0 | 0.74 | 0 |
| Critical Values | 1% | -3.43 | -3.43 | -3.43 | -3.43 |
|  | 5% | -2.86 | -2.86 | -2.86 | -2.86 |
|  | 10% | -2.56 | -2.56 | -2.56 | -2.56 |

Table.1 demonstrates that the ADF statistic for the 'Open' series is -1.09, even greater than the critical value at the 10% confidence level, and the P-value is significantly non-zero. Thus, we cannot reject the null hypothesis, suggesting that the 'Open' sequence can be considered non-stationary. The ADF statistic for the first-order differenced 'Open' sequence is -14.21, smaller than the critical value at the 1% confidence level, and the p-value is significantly 0. Hence, we can reject the null hypothesis, implying that the first-order differenced 'Open' sequence can be considered stationary.

From another perspective, we can verify our conclusion by plotting the line graphs of the 'Open' sequence, first-order differenced 'Open' sequence, 'Close' sequence, and first-order differenced 'Close' sequence as shown below. It's evident that for both 'Open' and 'Close', without differencing, the expected values significantly increase with time, indicating that the original sequences are unlikely to be stationary. However, after taking the first-order difference, the sequences fluctuate around 0 as time changes, showing no significant changes in expectation and variance. It reflects the characteristics of stationarity.

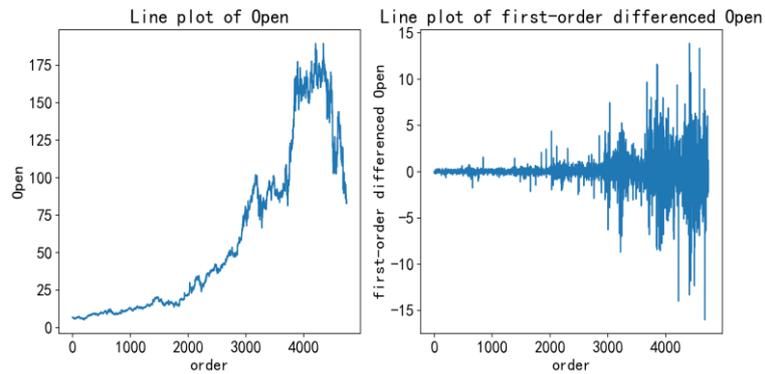

**Fig. 3.** Line Chart of Open sequence and First Order Differenced Open sequence.(Photo/Picture credit :Original)

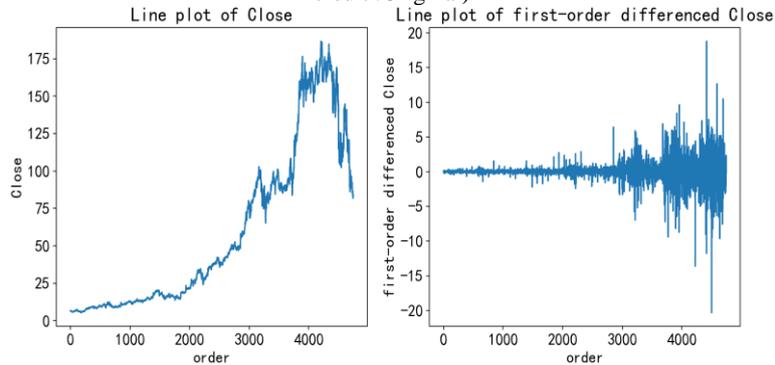

**Fig. 4.** Line Chart of Close sequence and First Order Differenced Close sequence. (Photo/Picture credit :Original)

### 2.5 Model Order Determination

Model order determination
First, we plot the autocorrelation function (ACF) and partial autocorrelation function (PACF) plots for the first-order differenced 'Open' sequence and the first-order differenced ' Close' sequence.

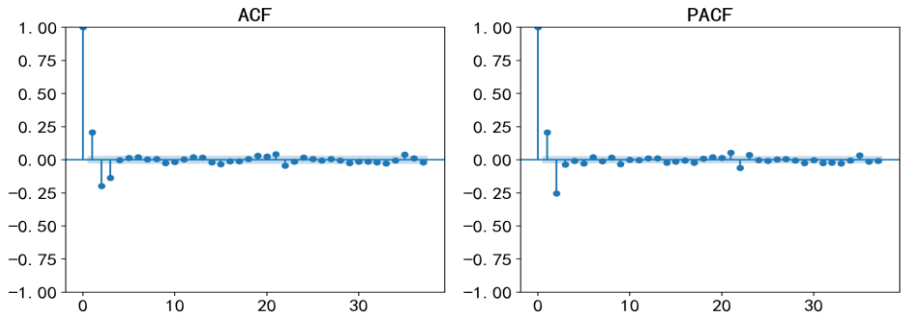

**Fig. 5.** the ACF and PACF plot for the first-order differenced 'Open' sequence. (Photo/Picture credit :Original)

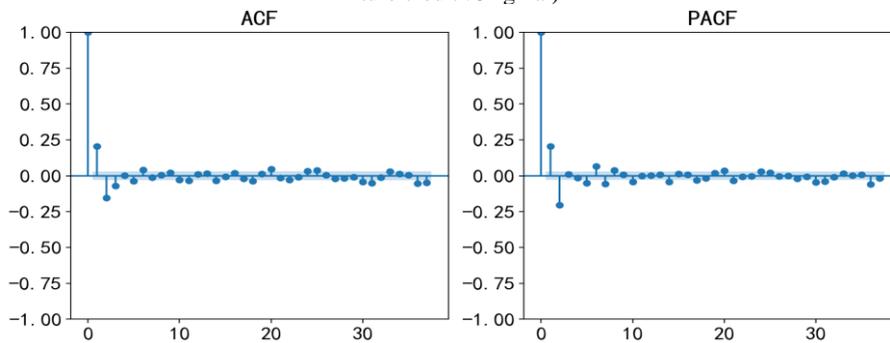

**Fig. 6.** the ACF and PACF plot for the first-order differenced 'Close' sequence. (Photo/Picture credit :Original)

Fig.5 implies that the ACF of Open sequence appears to have a cutoff at lag 2 or 3, and the PACF of Open sequence has a cutoff at lag 2. Similarly, Fig.6 implies that the ACF of Close sequence appears to have a cutoff at lag 2 or 3, and the PACF of Close sequence has a cutoff at lag 2.

Next, we generate the heatmaps for AIC and BIC, and make further judgments based on the AIC and BIC information criteria.

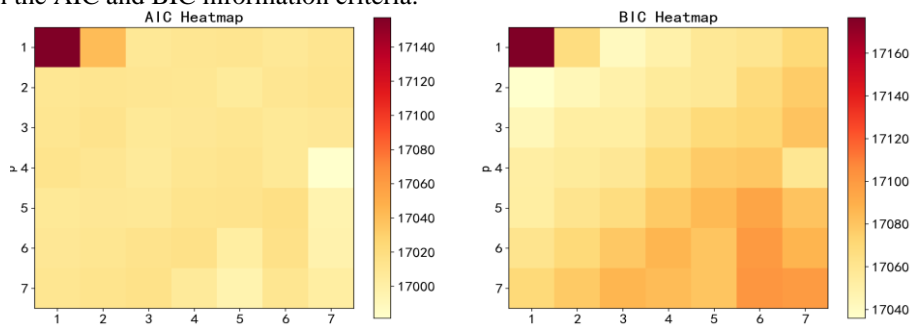

**Fig. 7.** the AIC Heatmap and BIC Heatmap of First Order Differenced Open. (Photo/Picture credit: Original)

Fig.7 indicates that the model order ARIMA(2,1,2) has a lower BIC compared to the model order ARIMA(2,1,3). Therefore, the ARIMA order for forecasting the 'Open' sequence is determined as ARIMA(2,1,2).

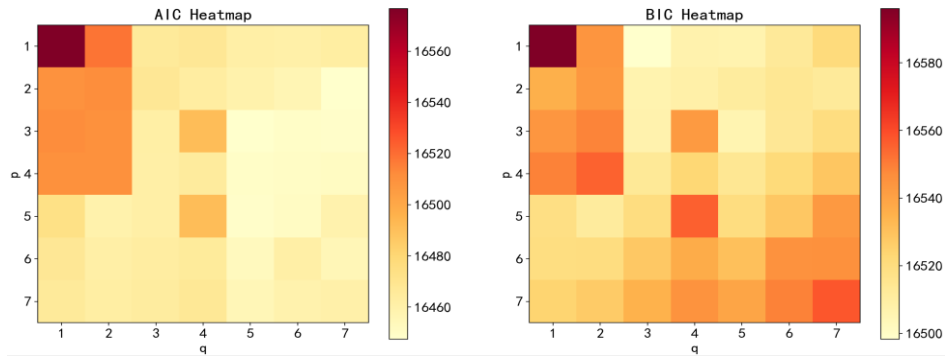

**Fig. 8.** the AIC Heatmap and BIC Heatmap of First Order Differenced Close.
(Photo/Picture credit: Original)

Fig.8 indicates that the model order ARIMA(2,1,3) has lower AIC and BIC values compared to the model order ARIMA(2,1,2). Therefore, the ARIMA order for forecasting the 'Close' sequence is determined as ARIMA(2,1,3).

### 2.6 Model Training and White Noise Test

**For the 'Open' sequence:.**
Based on the analysis above, we have determined the ARIMA order for the 'Open' sequence as ARIMA(2,1,2). We trained the model and conducted the Ljung-Box test for the residuals, resulting in the following Ljung-Box statistic (Lb) and p-value:

**Table. 2.** White Noise Test Results for ARIMA(2,1,2).

| Lag | Ljung-Box Statistic | p-value |
| --- | --- | --- |
| 1 | 0.000009 | 0.997600 |
| 2 | 0.000328 | 0.999836 |
| 3 | 0.170918 | 0.982142 |
| 4 | 1.149319 | 0.886370 |
| 5 | 4.386553 | 0.495204 |
| ⋮ | ⋮ | ⋮ |

Table.2 demonstrates that the p-values for all orders are greater than 0.05, so we fail to reject the null hypothesis. Hence, we consider the residuals of the Open prediction model's forecast values and observed values to be a white noise sequence. This indicates that the model has effectively learned the underlying patterns in the Open time series, thereby possessing certain application value.

From a visual standpoint, we can plot the line graph, density distribution plot, ACF plot, and PACF plot of the residuals as shown below:

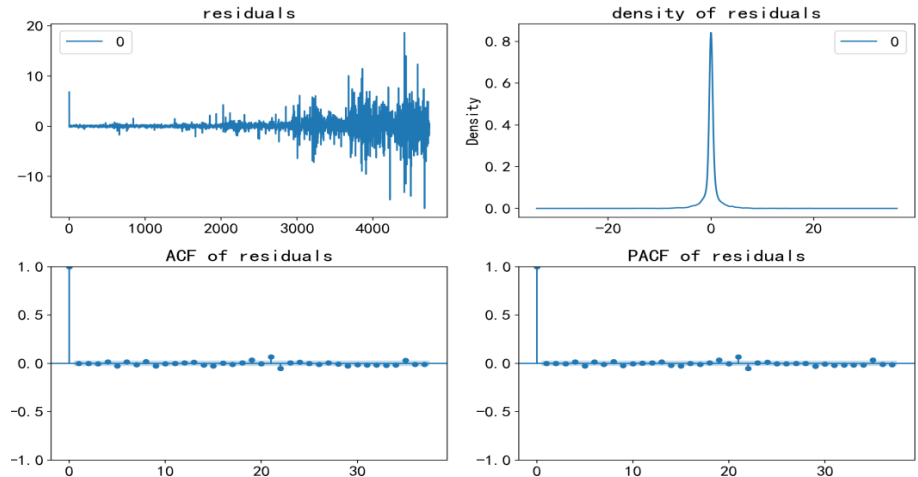

**Fig. 9.** Line plot, density distribution graph, ACF graph and PACF graph of residual series of Open series. (Photo/Picture credit :Original)

Fig.9 shows that the residual sequence fluctuates around 0, and it maintains symmetry above and below. The density distribution plot showcases a distribution of residuals that is close to a normal distribution. Analyzing the ACF and PACF plots, all autocorrelation coefficients and partial autocorrelation coefficients are close to 0, without any significant lags or spikes. These characteristics validate our conclusion.

**For the 'Close' sequence:**

Based on the analysis above, we have determined the ARIMA order for the 'Close' sequence as ARIMA(2,1,3). We trained the model and conducted the Ljung-Box test for the residuals, resulting in the following Ljung-Box statistic and p-value:

**Table. 3.** White Noise Test Results for ARIMA(2,1,3).

| Lag | Ljung-Box Statistic | p-value |
|---|---|---|
| 1 | 0.000198 | 0.988772 |
| 2 | 0.111173 | 0.945930 |
| 3 | 0.181199 | 0.980566 |
| 4 | 5.258114 | 0.261824 |
| 5 | 6.089130 | 0.297643 |
| ⋮ | ⋮ | ⋮ |

Table.3 demonstrates that p-values for all orders are greater than 0.05, so we fail to reject the null hypothesis. Hence, we consider the residuals of the Close prediction model's forecast values and observed values to be a white noise sequence. This indicates that the model has effectively learned the underlying patterns in the Close time series, thereby possessing certain application value.

From a visual standpoint, we can plot the line graph, density distribution plot, ACF plot, and PACF plot of the residuals as shown below:

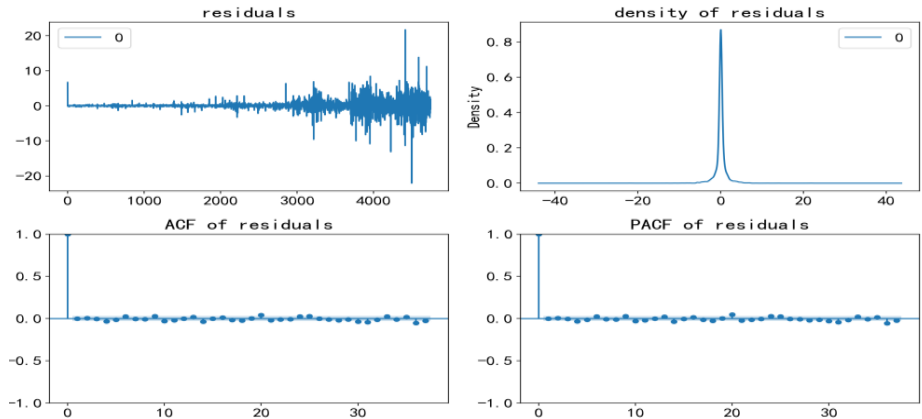

**Fig. 10.** Line plot; density distribution graph; ACF graph and PACF graph of residual series of Close series. (Photo/Picture credit :Original)

Fig.10 shows that the residual sequence fluctuates around 0, and it maintains symmetry above and below. The density distribution plot showcases a distribution of residuals that is close to a normal distribution. Analyzing the ACF and PACF plots, all autocorrelation coefficients and partial autocorrelation coefficients are close to 0, without any significant lags or spikes. These characteristics validate our conclusion.

### 2.7    Model Evaluation

The rolling window method is employed to partition the dataset for model validation. For a specific length of the test set, the rolling window initially contains data of that length. At the start of each iteration, the data within the rolling window is divided into a training set, and the next set of data of the same length is designated as the test set. The model is trained using the current training set and predictions are made. Finally, the model's mean squared error is calculated based on the predicted values and observed values of the test set to evaluate the model. Then, the rolling window is expanded to include the current test set, and this iteration ends. This process continues until the rolling window expands to its maximum size.

According to the above method, calculate the model's mean squared error for different test set lengths and create a line graph as shown below:

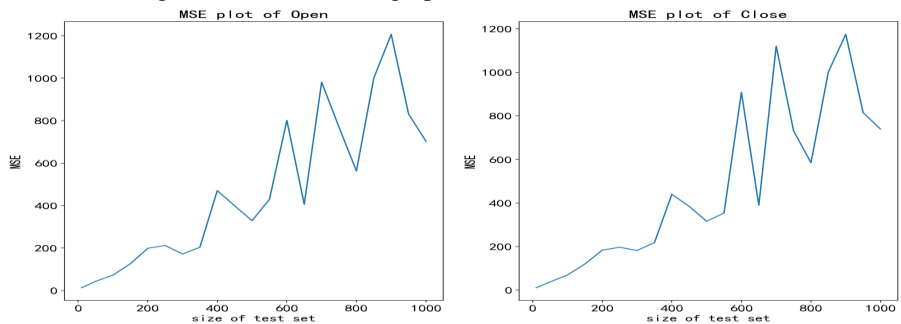

**Fig. 11.** Mean Squared Error plot of both models. (Photo/Picture credit :Original)

Fig.11 implies that the patterns exhibited by both the 'Open' sequence model and the 'Close' sequence model are the same. The mean squared error increases with an increase in the test set size. When the test set size is 10, the mean squared error is only around 10. However, when the test set size increases to 1000, the mean squared error rapidly increases to around 700. This reflects the characteristic of the ARIMA model, which has higher accuracy in forecasting short-term future data but is less effective in predicting long-term future data. Therefore, it is advisable to use the ARIMA model only for forecasting short-term future data.

## 3 Result

### 3.1 Forecasting the 'Open' Sequence

Establishing an Autoregressive Integrated Moving Average (ARIMA) model with orders (2,1,2) for the 'Open' sequence. All known data has been interpolated and input into the model for training. Using this model to forecast the 'Open' sequence data from January 1, 2023, to January 31, 2023.

Table.4 presents the model's fitted coefficients and relevant statistics such as AIC, BIC, standard deviation, Z-score, p-value, 95% confidence interval, etc.

is as follows:

**Table. 4.** Fitting Coefficients and Correlation Statistics of the Stock Opening Price Forecasting Model.

| Model  |        |         | ARIMA(2,1,2) |       |        |        |
|--------|--------|---------|--------------|-------|--------|--------|
| AIC    |        |         | 17011.601    |       |        |        |
| BIC    |        |         | 17043.923    |       |        |        |
| HQIC   |        |         | 17022.960    |       |        |        |
|        | coef   | std err | z            | P>\|z\| | [0.025 | 0.975] |
| ar.L1  | 0.4200 | 0.032   | 13.002       | 0.000 | 0.357  | 0.483  |
| ar.L2  | -0.2575| 0.027   | -9.595       | 0.000 | -0.310 | -0.205 |
| ma.L1  | -0.1707| 0.034   | -4.998       | 0.000 | -0.238 | -0.104 |
| ma.L2  | -0.0319| 0.032   | -0.996       | 0.319 | -0.095 | 0.031  |
| sigma2 | 2.1099 | 0.013   | 159.876      | 0.000 | 2.084  | 2.136  |

### 3.2 Forecasting the 'Close' Sequence .

Establishing an Autoregressive Integrated Moving Average (ARIMA) model with orders (2,1,3) for the 'Close' sequence. All known data has been interpolated and input into the model for training. Using this model to forecast the 'Close' sequence data from January 1, 2023, to January 31, 2023.

Table.5 presents the model's fitted coefficients and relevant statistics such as AIC, BIC, standard deviation, Z-score, p-value, 95% confidence interval, etc.

Table. 5. Fitting Coefficients and Correlation Statistics of the Stock Closing Price Forecasting Model.

| Model | | ARIMA(2,1,3) | | | |
|---|---|---|---|---|---|
| AIC | | 16467.892 | | | |
| BIC | | 16506.679 | | | |
| HQIC | | 16522.677 | | | |
| | coef | std err | z | P>|z| | [0.025  0.975] |
| ar.L1 | -0.6376 | 0.041 | -15.708 | 0.000 | -0.717  -0.558 |
| ar.L2 | -0.0301 | 0.043 | -0.697 | 0.486 | -0.115  0.054 |
| ma.L1 | 0.8957 | 0.041 | 22.086 | 0.000 | 0.816  0.975 |
| ma.L2 | 0.0384 | 0.053 | 0.722 | 0.470 | -0.066  0.143 |
| ma.L3 | -0.1787 | 0.015 | -11.955 | 0.000 | -0.208  -0.149 |
| sigma2 | 1.8806 | 0.011 | 174.229 | 0.000 | 1.859  1.902 |

### 3.3 Display of Fitting Effects for Both Model.

From January 1, 2010, to December 31, 2022, data points were taken at intervals of 100 days. The observed values-time line graph and predicted values-time line graph for both the 'Open' and 'Close' sequences are shown below in black and blue respectively. It can be observed that the constructed ARIMA models achieve accurate predictions for the majority of the time periods .

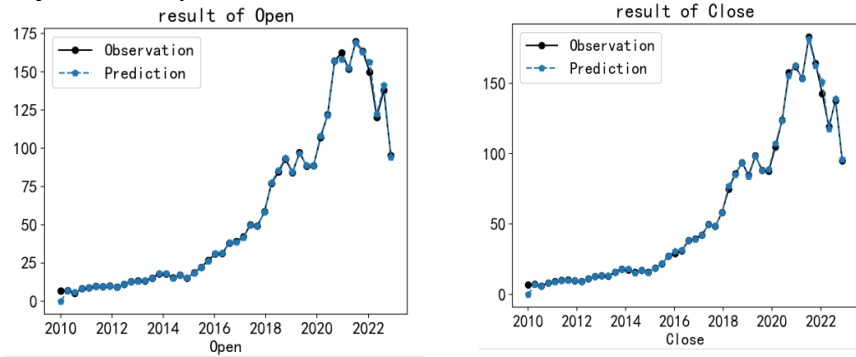

**Fig. 12.** Time-Observed Values Plot and Time-Forecasted Values plot for Two Models.
(Photo/Picture credit: Original)

## 4 Conclusion

To address the challenge posed by predicting time series with missing values that disrupt sequence continuity, this paper proposes a short-term time series forecasting model based on the fusion of cubic spline interpolation and the ARIMA algorithm. The methodology involves initially applying cubic spline interpolation to the time series with missing values. Subsequently, a self-regressive integrated moving average (ARIMA) model is constructed using the interpolated and smoother time series, allowing for the prediction of short-term time series values for the future. The short-term time series forecasting model proposed in this paper exhibits a straightforward and easily interpretable principle. It demonstrates excellent fitting performance, and the model's predictive results hold certain reference value for the short-term period. However, the model also has its limitations. Firstly, the model is inherently suitable only for short-term predictions, and its accuracy diminishes for longer-term continuous predictions. Secondly, in cases where the sequence exhibits intense fluctuations, the predictive accuracy of the model is compromised. In the forthcoming research, efforts will be directed towards addressing these shortcomings, aiming to modify and optimize the model to explore the optimal approaches for forecasting stock time series.

# Reference

1. Xiaoli Zhao, F.: Application Research of ARIMA Model in Short-Term Passenger Flow Forecasting for Urban Rail Transit. Modern Urban Transit, 77-82(2023).
2. Zhuang Liu, F.: Prediction of Chlorophyll-a Concentration in Changtan Reservoir Based on ARIMA Model. Environmental Pollution and Control 45 (7),895-896(2023).
3. Xiao Wang, F.: Effect of Oceanic Niño index on interannual CPUE of yellowfin tuna (Thunnus albacares) in Western and Central Pacific Ocean based on ARIMA model. South China Fishers Science 19 (4),11-13(2023).
4. Huihui Wu, F.: Global Surface Temperature Forecast Analysis Based on ARIMA Model. Modern Information Technology 7(16),147-150(2023).
5. Jian He, F., Xiaofang Wang, S.: Short-Term Wind Speed Forecast Based on Combined ARIMA and LS-SVM Model. Mechanical and Electrical Engineering Technology 52(8),30-33(2023).
6. Yucui Li, F.: Analysis of Development Trends in Sino-Vietnamese Cross-Border E-Commerce Based on ARIMA Model. Foreign Trade and Economic Cooperation 34(9),6-10(2023).
7. Jianguo Jiang, F.: Forecast and Analysis of Pulmonary Tuberculosis Epidemic Trends in Henan Province Using the ARIMA Model. Modern Disease Control and Prevention 34(7),495-498(2023).
8. Long Gao, F.: Research on Small-Batch Material Production Demand Forecasting Model Based on ARIMA. Modern Information Technology7(15), 100-101(2023).
9. Lujun Xu, F.: Forecasting Enterprise Free Cash Flow Based on ARIMA Model.Financial observation,47-48 (2023).
10. Lijuan Wang, F. Hui Suo, F.: Usage and Analysis of Interpolation Methods in Time Series Forecasting. Software development and application,13-15.(2022).